\documentclass[runningheads]{llncs}
\usepackage[T1]{fontenc}
\usepackage{graphicx}
\usepackage{amsmath}
\usepackage{multirow}
\usepackage{booktabs}
\usepackage{hyperref}
\usepackage{cite}
\usepackage{color}

\begin{document}
\title{Accent-VITS: accent transfer for end-to-end TTS}
%
%

\author{Linhan Ma \inst{1} \and
Yongmao Zhang\inst{1} \and
Xinfa Zhu\inst{1} \and
Yi Lei\inst{1} \and
Ziqian Ning\inst{1} \and
Pengcheng Zhu\inst{2} \and
Lei Xie\inst{1}\thanks{Corresponding author.} }
\authorrunning{L. Ma et al.}
%
\institute{Audio, Speech and Language Processing Group (ASLP@NPU), School of Computer Science, Northwestern Polytechnical University, Xi'an, China \and
Fuxi AI Lab, NetEase Inc., Hangzhou, China \\
}
\maketitle              

\begin{abstract}
Accent transfer aims to transfer an accent from a source speaker to synthetic speech in the target speaker's voice. The main challenge is how to effectively disentangle speaker timbre and accent which are entangled in speech. This paper presents a VITS-based ~\cite{vits} end-to-end accent transfer model named \textit{Accent-VITS}. 
Based on the main structure of VITS, Accent-VITS makes substantial improvements to enable effective and stable accent transfer. 
We leverage a hierarchical CVAE structure to model accent pronunciation information and acoustic features, respectively, using bottleneck features and mel spectrums as constraints.
Moreover, the text-to-wave mapping in VITS is decomposed into text-to-accent and accent-to-wave mappings in Accent-VITS.
In this way, the disentanglement of accent and speaker timbre becomes be more stable and effective. 
Experiments on multi-accent and Mandarin datasets show that Accent-VITS achieves higher speaker similarity, accent similarity and speech naturalness as compared with a strong baseline\footnote{Demos: \url{https://anonymous-accentvits.github.io/AccentVITS/}}.

\keywords{Text to speech  \and Accent transfer \and Variational autoencoder \and Hierarchical.}
\end{abstract}
\section{Introduction}
In recent years, there have been significant advancements in neural text-to-speech (TTS), which can generate human-like natural speech from input text. Accented speech is highly desired for a better user experience in many TTS applications. Cross-speaker accent transfer is a promising technology for accented speech synthesis, which aims to transfer an accent from a source speaker to the synthetic speech in the target speaker's voice. Accent transfer can promote cross-region communication and make a TTS system better adapt to diverse language environments and user needs. 

An accent is usually reflected in the phoneme pronunciation pattern and prosody variations, both of which are key attributes of the accent rendering \cite{LootsN11, MareuilV06, DBLP:journals/corr/abs-2209-10804}. 
The segmental and suprasegmental structures may be in distinctive pronunciation patterns for different accents and influence the listening perception of speaking accents \cite{KolluruWLYG14, abs-2305-04816}. 
The prosody variations of accent are characterized by different pitch, energy, duration, and other prosodic appearance. To build an accent TTS system using accent transfer, the research problem can be treated as how to effectively \textit{disentangle} speaker timbre and accent factors in speech. However, it is difficult to force the system to sufficiently disentangle the accent from the speaker timbre and content in speech since both pronunciation and prosody attributes are featured by local variations at the fine-grained level.
And usually, each speaker has only one accent in the training phase which adds to the difficulty of disentangling.

Previous approaches attempting to disentangle accent attributes and speaker timbre are mainly based on Domain Adversarial Training (DAT) ~\cite{DAT}. However, when the feature extraction function has a high capacity, DAT poses a weak constraint to the feature extraction function. Therefore, a single classifier with a gradient reversal layer in accent transfer TTS cannot disentangle the accent from the speaker's timbre, as the accent is varied in prosody and pronunciation. Additionally, gradient descent in domain adversarial training can violate the optimizer's asymptotic convergence guarantees, often hindering the transfer performance ~\cite{AcunaLZF22}. Applying DAT in accent transfer tasks, especially when each speaker has only one accent in the training phase, may result in inefficient and unstable feature disentanglement. Furthermore, there is a trade-off between speaker similarity and accent similarity, which means entirely removing speaker timbre hurts performance on preserving accent pronunciation ~\cite{ShuBNE18}.

Bottleneck (BN) features are recently used as an intermediate representation to supervise accent attribute modeling in TTS ~\cite{ZhangWYSWX22}. The BN feature, extracted from a well-trained neural ASR model, is considered to be noise-robust and speaker-independent \cite{bnrecog, BNrobust}, which benifits speaker timbre and accent disentanglement. 
However, in the methods with BN as an intermediate representation \cite{bnTTS1, Hiertron}, models are often trained independently in multiple stages.
This can lead to the issue of error accumulation and model mismatch between each stage, resulting in the degradation of synthesized speech quality and accent attributes.


In this paper, we propose an end-to-end accent transfer model, \textit{Accent-VITS}, with a hierarchical conditional variational autoencoder (CVAE) ~\cite{HierSpeech} utilizing bottleneck features as a constraint to eliminate speaker timbre from the original signal.
Specifically, we leverage the end-to-end speech synthesis framework, VITS ~\cite{vits}, as the backbone of our model, since it achieves good audio quality and alleviates the error accumulation caused by the conventional two-stage TTS system consisting of an acoustic model and a vocoder.
Based on the VITS structure, an additional CVAE is added to extract an accent-dependent latent distribution from the BN feature. The latent representation contains the accent and linguistic content and is modeled by the accented phoneme sequence input.
The BN constraint factorizes the cross-speaker accent TTS into two joint-training processes, which are text-to-accent and accent-to-wave. The \textit{text-to-accent} process takes the accented phoneme sequence as input to generate an accent-dependent distribution. The \textit{accent-to-wave} process produces the speech distribution in the target accent and target speaker from the output accent distribution and is conditioned on speaker identity. This design enables more effective learning of accent attributes, leading to sufficient disentanglement and superior performance of accent transfer for the synthesized speech.
Experimental results on Mandarin multi-accent datasets demonstrate the superiority of our proposed model.

\begin{figure}
\includegraphics[width=\textwidth]{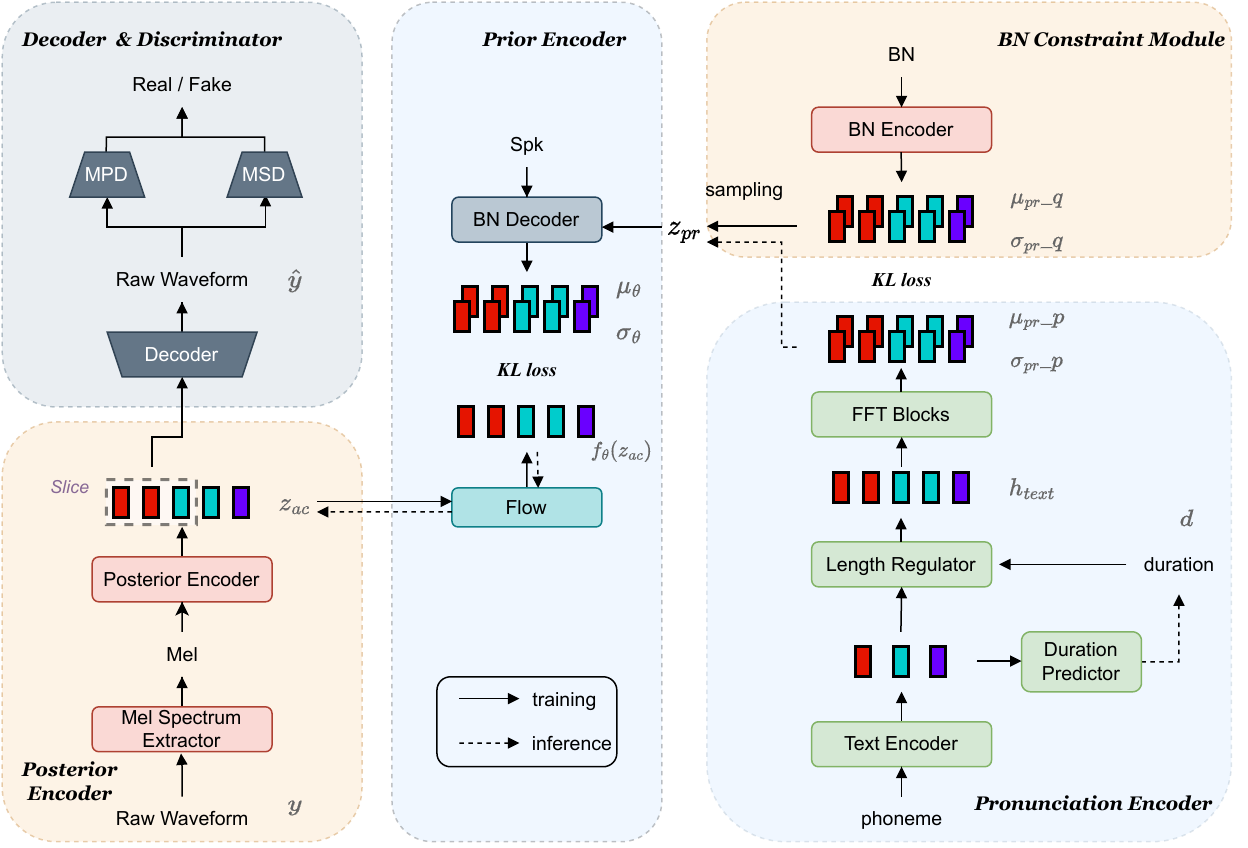}
\caption{Overview of Accent-VITS structure.} \label{fig:model}
\end{figure}

\section{Method}

This paper proposes a VITS-based end-to-end accent transfer model with a hierarchical conditional variational autoencoder (CVAE). As shown in Fig. \ref{fig:model}, it mainly includes five parts: a posterior encoder, a decoder, a prior encoder, a pronunciation encoder, and a BN constraint module.
The posterior encoder extracts the latent representation $z_{ac}$ of acoustic feature from the waveform $y$, and the decoder reconstructs the waveform $\hat{y}$ according to $z_{ac}$:
\begin{equation}
    z_{ac} = \text{PostEnc}(y) \sim q(z_{ac}|y) 
\end{equation}
\begin{equation}
    \hat{y} = \text{Dec}(z_{ac}) \sim p(y|z_{ac})
\end{equation}
The prior encoder produces a prior distribution of $z_{ac}$. Since the prior encoder of CVAE in VITS cannot effectively disentangle accent from the speaker timbre and text content, we use a hierarchical CVAE structure to model accent information and acoustic features sequentially. We take the bottleneck feature (BN) that is extracted from the source wav by an ASR system as a constraint to improve the pronunciation information of accent in the latent space. The BN Encoder extracts the latent representation $z_{pr}$ which contains accent pronunciation information from BN. The pronunciation encoder gets a prior distribution $p(z_{pr}|c)$ of the latent variables $z_{pr}$ given accented phoneme sequence condition $c$.
The BN decoder in the prior encoder module also acts as the decoder of the first CVAE structure, getting the prior distribution $p(z_{ac}|z_{pr}, spk)$ of the latent variables $z_{ac}$ from the 
sampled latent representation $z_{pr}$ given speaker identity condition $spk$.
This hierarchical CVAE adopts a reconstruction objective $L_{recon}$ and two prior regularization terms as
\begin{equation}
    \begin{aligned}
        L_{cvae} = \alpha L_{recon} + D_{KL}(q(z_{pr}|BN)||p(z_{pr}|c)) + &\\ 
        D_{KL}(q(z_{ac}|y)||p(z_{ac}|z_{pr},spk&))
    \end{aligned}
\end{equation}
where $D_{KL}$ is the Kullback-Leibler divergence.
For the reconstruction loss, we use L1 distance of the mel-spectrum between ground truth and generated waveform. In the following, we will introduce the details of these modules.

\subsection{Pronunciation Encoder}
We assign a different phoneme set to each accent (standard Mandarin or accent Mandarin in this paper) and use a rule-based converter (G2P tool) to get the phoneme sequence from the text. Given the phoneme sequence of accent or Mandarin condition $c$, the pronunciation encoder module predicts the prior distribution $p(z_{pr}|c)$ used for the prior regularization term of the first CVAE structure. In this module, the text encoder which consists of multiple FFT ~\cite{transformer} blocks takes phoneme sequences as input and produces phoneme-level representation.
Different from VITS, we use the length regulator (LR) in FastSpeech ~\cite{fs} to extend the phoneme-level representation to frame-level representation $h_{text}$ ~\cite{VISinger}. 
The other multiple FFT blocks are used to extract a sequence of hidden vectors from the frame-level representation $h_{text}$ and then generate the mean $\mu_{pr}\_p$ and variance $\sigma_{pr}\_p$ of the prior normal distribution of latent variable $z_{pr}$ by a linear projection.
\begin{equation}
    p(z_{pr}|c) = N(z_{pr}; \mu_{pr}\_p(c),  \sigma_{pr}\_p(c))
\end{equation}

\subsection{BN Constraint Module}
In this module, the BN encoder extracts the latent representation of pronunciation information $z_{pr}$ from the BN feature and produces the posterior normal distribution $q(z_{pr}|BN)$ with the mean $\mu_{pr}\_q$ and variance $\sigma_{pr}\_q$. BN feature is usually the feature map of a neural network layer. Specifically, the BN adopted in this paper is the output of an ASR encoder, which is generally considered to contain only linguistic and prosodic information such as pronunciation, intonation, accent, and very limited speaker information ~\cite{BNrobust}. The ASR model is usually trained with a large multi-speaker multi-condition dataset, and the BN feature extracted by it is also believed to be noise-robust and speaker-independent. The BN encoder consists of multiple layers of Conv1d, ReLU activation, Layer Normalization, Dropout, and a layer of linear projection to produce the mean and variance.

\subsection{Prior Encoder}
The BN decoder in the prior encoder module also acts as the decoder of the first CVAE structure. 
Given the speaker identity condition $spk$, the BN decoder extracts the latent representation of acoustic feature from sampled $z_{pr}$ and generates the prior normal distribution with mean $\mu_{\theta}$ and variance $\sigma_{\theta}$ of $z_{ac}$.
Following VITS, a normalizing flow \cite{flowstructure, flowVae} $f_\theta$ is added to the prior encoder to improve the expressiveness of the prior distribution of the latent variable $z_{ac}$.
\begin{equation}
    \begin{aligned}
        p(f_\theta(z_{ac})|z_{pr},spk) = N(f_\theta(z_{ac}); \mu_\theta(z_{pr},spk), \sigma_\theta(z_{pr},spk))  
    \end{aligned}
\end{equation}
\begin{equation}
    \begin{aligned}
        p(z_{ac}|z_{pr},spk) = p(f_\theta(z_{ac})|z_{pr},spk)\left | \det \frac{\partial f_\theta(z_{ac})}{\partial z_{ac}}\right |
    \end{aligned}
\end{equation}

\subsection{Posterior Encoder}

The posterior encoder module extracts the latent representation $z_{ac}$ from the waveform $y$. The mel spectrum extractor in it is a fixed signal processing layer without updatable weights.
The encoder firstly extracts the mel spectrum from the raw waveform through the signal processing layer.
Unlike VITS, the posterior encoder takes the mel spectrum as input instead of the linear spectrum. We use multiple layers of Conv1d, ReLU activation, Layer Normalization, and Dropout to extract a sequence of hidden vector and then produces the mean and variance of the posterior distribution $q(z_{ac}|y)$ by a Conv1d layer. Then we can get the latent $z_{ac}$ sampled from $q(z_{ac}|y)$ using the reparametrization trick.

\subsection{Decoder}

The decoder generates audio waveforms from the intermediate representation $z_{ac}$. We use HiFi-GAN generator G ~\cite{HifiGAN} as the decoder. For more efficient training, we only feed the sliced $z_{ac}$ instead of the entire length into the decoder to generate the corresponding audio segment. We also use GAN-based ~\cite{GANpaper} training to improve the quality of the synthesized speech. The discriminator D follows HiFiGAN’s Multi-Period Discriminator (MPD) and Multi-Scale Discriminator (MSD) ~\cite{HifiGAN}. Specifically, the GAN losses for the generator G and discriminator D are defined as:
\begin{equation}
    \begin{aligned}
        L_{adv}(G) = E_{(z_{ac})} \left [ (D(G(z_{ac}))-1)^{2} \right ]
    \end{aligned}
\end{equation}
\begin{equation}
    \begin{aligned}
        L_{adv}(D) = E_{(y,z_{ac})}\left [ (D(y)-1)^{2}+(D(G(z_{ac})))^{2}\right ]
    \end{aligned}
\end{equation}

\begin{table*}[]
\renewcommand\arraystretch{1.2}
\caption{Experimental results in terms of subjective mean opinion score (MOS) with confidence intervals of 95\% and two objective metrics. Note: the results of VITS-DAT are missing as the system cannot converge properly during training.}
\label{expresults}
\begin{tabular}{ccccc|cc}
\toprule[1pt]
\multirow{4}{*}{Accent} & \multirow{4}{*}{Model} & \multicolumn{3}{c|}{\multirow{2}{*}{Subjective Evaluation}}                                                                                  & \multicolumn{2}{c}{\multirow{2}{*}{Objective Evaluation}}                             \\
                        &   & \multicolumn{3}{c|}{}     & \multicolumn{2}{c}{}      \\ \cline{3-7} 
                        &                        & \multirow{2}{*}{SMOS$\uparrow$}                          & \multirow{2}{*}{NMOS$\uparrow$}                          & \multirow{2}{*}{AMOS$\uparrow$}                          & Speaker cosine  & Duration       \\
                        &   &   &   &   & {Similarity$\uparrow$} &{MAE$\downarrow$} \\
\midrule[1pt]
\multirow{2}{*}{Shanghai}                & T2B2M                  & 3.85$\pm$0.03                           & 3.68$\pm$0.06                           & 3.73$\pm$0.05                           & 0.76                                & 3.51                           \\
                        & Accent-VITS            & \textbf{3.86$\pm$0.02} & \textbf{3.76$\pm$0.02} & \textbf{3.76$\pm$0.06} & \textbf{0.83}      & \textbf{3.06} \\ \hline
\multirow{2}{*}{Henan}                   & T2B2M                  & 3.79$\pm$0.07                           & 3.75$\pm$0.03                           & 3.59$\pm$0.02                           & 0.78                                & 3.65                           \\
                        & Accent-VITS            & \textbf{3.87$\pm$0.02} & \textbf{3.92$\pm$0.04} & \textbf{3.62$\pm$0.04} & \textbf{0.81}      & \textbf{3.21} \\ \hline
\multirow{2}{*}{Dongbei}                 & T2B2M                  & 3.88$\pm$0.05                           & 3.87$\pm$0.03                           & 3.50$\pm$0.03                           & 0.79                                & 3.49                           \\
                        & Accent-VITS            & \textbf{4.01$\pm$0.02} & \textbf{4.14$\pm$0.05} & \textbf{3.88$\pm$0.02} & \textbf{0.87}      & \textbf{3.01} \\ \hline
\multirow{2}{*}{Sichuan}                 & T2B2M                  & 3.94$\pm$0.06                           & \textbf{3.82$\pm$0.04} & 3.69$\pm$0.06           & \textbf{0.84}      & 3.55                           \\
                        & Accent-VITS            & \textbf{4.06$\pm$0.04} & 3.80$\pm$0.02  & \textbf{3.81$\pm$0.06} & \textbf{0.84}      & \textbf{3.14} \\ \hline
\multirow{2}{*}{Average}                 & T2B2M                  & 3.87$\pm$0.05                           & 3.78$\pm$0.04                           & 3.63$\pm$0.02                           & 0.79                                & 3.55                           \\
                        & Accent-VITS            & \textbf{3.95$\pm$0.04} & \textbf{3.91$\pm$0.03} & \textbf{3.77$\pm$0.06} & \textbf{0.84}      & \textbf{3.11} \\ 
\bottomrule[1pt]
\end{tabular}
\end{table*}

\subsection{Duration Predictor}
In the training process, the LR module expands the phoneme-level representation using the ground truth duration, denoted as $d$. In the inference process, the LR module expands the representation using the predicted duration, denoted as $\hat{d}$, obtained from the duration predictor. 
Unlike VITS, we utilize a duration predictor consisting of multiple layers of Conv1d, ReLU activation, Layer Normalization, and Dropout instead of a stochastic duration predictor due to the significant correlation between accent-specific pronunciation prosody and duration information ~\cite{VISinger}. For the duration loss $L_{dur}$, we use MSE loss between $\hat{d}$ and $d$.

\subsection{Final Loss}
With the above hierarchical CVAE and adversarial training, we optimize our proposed model with the full objective:
\begin{equation}
    \begin{aligned}
        L = L_{adv}(G) + L_{fm}(G) + L_{cvae} + \lambda L_{dur}
    \end{aligned}
\end{equation}
\begin{equation}
    \begin{aligned}
        L(D) = L_{adv}(D)
    \end{aligned}
\end{equation}
where $L_{adv}(G)$ and $L_{adv}(D)$ are the GAN loss of G and D respectively, and feature matching loss $L_{fm}$ is added to improve the stability of the training. The $L_{cvae}$ consists of the reconstruction loss and two KL losses.

\section{Experiments}

\subsection{Datasets}
The experimental data consists of high-quality standard Mandarin speech data and accent Mandarin speech data from four different regions: Sichuan, Dongbei (Northeast China), Henan, and Shanghai. Specifically, we use DB1\footnote{\url{https://www.data-baker.com/open_source.html}} as the high-quality standard Mandarin data which contains 10,000 utterances recorded in a studio from a professional female anchor. The total duration is approximately 10.3 hours.
The accent data from the four regions were also recorded in a recording studio by speakers from these regions. 
Among them, Sichuan, Dongbei, and Shanghai each have two speakers, one male and one female respectively, while Henan has only one female speaker.
In detail, Sichuan, Dongbei, Henan, and Shanghai accent data have 2794, 3947, 2049, and 4000 utterances, respectively. The duration of the accent data is approximately 13.7 hours. So we have a total of 4 accents and 8 speakers in our training data.

All the audio recordings are downsampled to 16kHz. We utilize 80-dim mel-spectrograms with 50ms frame length and 12.5ms frame shift. Our ASR model is based on the WeNet U2++ model ~\cite{wenet} trained on 10,000 hours of data from the WenetSpeech corpus ~\cite{wenetspeech}. We use the Conformer-based encoder output as our BN feature with 512-dim. The BN feature is further interpolated to match the sequence length of the mel-spectrogram.

\begin{table*}[]
\renewcommand\arraystretch{1.2}
\caption{Results of ablation studies.}
\label{ablation}
\begin{tabular}{ccccc|cc}
\toprule[1pt]
\multirow{4}{*}{Accent} & \multirow{4}{*}{Model} & \multicolumn{3}{c|}{\multirow{2}{*}{Subjective Evaluation}}                                                                                  & \multicolumn{2}{c}{\multirow{2}{*}{Objective Evaluation}}                             \\
                        &   & \multicolumn{3}{c|}{}     & \multicolumn{2}{c}{}      \\ \cline{3-7} 
                        &                        & \multirow{2}{*}{SMOS$\uparrow$}                          & \multirow{2}{*}{NMOS$\uparrow$}                          & \multirow{2}{*}{AMOS$\uparrow$}                          & Speaker cosine  & Duration       \\
                        &   &   &   &   & {Similarity$\uparrow$} &{MAE$\downarrow$} \\
\midrule[1pt]
\multirow{4}{*}{Average} & Accent-VITS            & \textbf{3.95$\pm$0.04} & \textbf{3.91$\pm$0.03} & \textbf{3.77$\pm$0.06} & \textbf{0.84}      & \textbf{3.11} \\
                         & -BN encoder            & 3.82$\pm$0.04                           & 3.72$\pm$0.01                           & 3.61$\pm$0.02                           & 0.77                                & 3.41                           \\
                         & -BN decoder            & 3.89$\pm$0.06                           & 3.86$\pm$0.03                           & 3.66$\pm$0.03                           & 0.82                                & 3.13                           \\
                         & -BN (enc,dec)          & 3.05$\pm$0.07                           & 3.24$\pm$0.04                           & 2.93$\pm$0.07                           & 0.69                                & 4.03                           \\ 
\bottomrule[1pt]
\end{tabular}
\end{table*}

\subsection{Model Configuration}
We implemented the following three models for comparison.
\begin{itemize}
    \item [$\bullet$]Text2BN2Mel (T2B2M) \cite{ZhangWYSWX22, Hiertron}: a three-stage accent transfer system composed of independently trained models for Text2BN, BN2Mel, and neural Vocoder. The Text2BN model predicts BN feature that contains accent pronunciation and content information from the input text. The BN2Mel model predicts mel-spectrogram based on the input BN feature and speaker identity. HiFiGAN V1 is used as the vocoder.

    \item [$\bullet$]VITS-DAT: an accent transfer model based on VITS and DAT. For disentangling accent and speaker timbre information, we add a DAT module composed of a gradient reversal layer and a speaker classifier to the output of the text encoder in VITS. Note that we also use a non-stochastic duration predictor and LR module instead of a stochastic duration predictor and MAS ~\cite{vits} in this model.

    \item [$\bullet$]Accent-VITS: the proposed accent transfer model in this paper.

\end{itemize}

In the text frontend processing, we assigned a different phoneme set for each accent or standard Mandarin.
The ground truth phoneme-level duration of all datasets is extracted by force-alignment tools.
The above comparison models are trained for 400k steps. The batch size of all the models is 24. The initial learning rate of all the models is 2e-4. The Adam optimizer with $\beta1$ = 0.8, $\beta2$ = 0.99 and $\epsilon$ = $10^{-9}$ is used to train all them. 

\subsection{Subjective Evaluation}

The TTS test set consists of both short and long total of 30 sentences for each accent without overlap with the training set. 
Each test sentence was synthesized by combining all the speakers in the training data separately.
The VITS-DAT system can not converge due to the instability of the combination of variational inference and DAT. Therefore we do not evaluate the results of VITS-DAT here.
We randomly selected 20 synthetic utterances for each accent, resulting in a total of 80 utterances for subjective listening. We asked fifteen listeners for each accent to assess speaker similarity (SMOS) and speech naturalness (NMOS). There are twenty local accent listeners to evaluate the accent similarity (AMOS), with five listeners for each accent. 
Particularly for the SMOS test, we use target speakers’ real recordings as reference.
The results are summarized in Table \ref{expresults}.

\subsubsection{Speaker Similarity.} 
The results shown in Table \ref{expresults} indicate that Accent-VITS can achieve the best performance in speaker similarity. 
Among the transfer results of all four accents, the SMOS score of Accent-VITS is better than that of T2B2M. 
This shows that our end-to-end model Accent-VITS effectively avoids the error accumulation and mismatch problems in the multi-stage model so that it can synthesize speech with more realistic target speaker timbre.

\subsubsection{Speech Naturalness.} 
In the NMOS test, we ask the listeners to pay more attention to the general prosody such as rhythm and expressiveness of the audio.
From Table \ref{expresults} we can see that the NMOS score of Accent-VITS is very close to T2B2M on the Sichuan accent and outperforms T2B2M on the other three accents. This indicates that the speech synthesized by the end-to-end model Accent-VITS is more natural than T2B2M on average.


\subsubsection{Accent Similarity.} 
In the AMOS test, we ask accent listeners to assess the similarity between synthesized speech and target accent, ignoring the naturalness of general prosody. 
The results in Table \ref{expresults} show that Accent-VITS achieves a higher AMOS score than T2B2M, which indicates that Accent-VITS can model accent attribute information better than T2B2M thanks to its hierarchical modeling of accent information and acoustic features.

\subsection{Objective Evaluation}
Objective metrics, including speaker cosine similarity and duration mean absolute error (Duration MAE), are also calculated.

\subsubsection{Speaker Cosine Similarity.} 
We calculate the cosine similarity on the generated samples to further verify the speaker similarity. Specifically, we train an ECAPA-TDNN model ~\cite{cosmodel} using 6000 hours of speech from 18083 speakers to extract x-vectors. The cosine similarity to the target speaker audio is measured on all synthetic utterances. The results are also shown in Table \ref{expresults}. 
The speaker cosine similarity score of Accent-VITS is also higher than that of T2B2M on three accents except for the Sichuan accent.
Compared with the T2B2M, Accent-VITS gets higher scores of speaker cosine similarity in Shanghai, Henan, and Dongbei accents. And in the Sichuan accent, both are equal.
This further demonstrates that Accent-VITS is better than T2B2M in modeling the target speaker timbre.

\subsubsection{Duration MAE.} 
Prosody variations are key attributes of accent rendering, which is largely reflected in the perceived duration of pronunciation units. Therefore, we further calculate the duration mean absolute error between the predicted duration results of different models and the ground truth. 
The results in Table \ref{expresults} show that Accent-VITS gets lower Duration MAE scores than T2B2M in transfer results of all four accents, which means that the transfer results of Accent-VITS are closer to the target accent in prosody than the transfer results of T2B2M.

\subsection{Ablation Study}
To investigate the importance of our proposed methods in Accent-VITS, three ablation systems were obtained by dropping the BN encoder and BN decoder respectively, and dropping both of them simultaneously, referred to as \textit{-BN encoder}, \textit{-BN decoder}, and \textit{-BN (enc, dec)}.
When dropping the BN encoder alone, the FFT blocks module directly predicts BN as an intermediate representation. We use MSE loss between the predicted BN and the ground truth BN as the constraint. The BN decoder module takes BN as input.
When dropping the BN decoder alone, the distribution of $z_{pr}$ directly as the prior distribution of $z_{ac}$. The flow module takes the sampled $z_{pr}$ as input.
When dropping both of them simultaneously, the FFT blocks module predicts the distribution of BN as the prior distribution of $z_{ac}$. We use MSE loss between the sampled BN and the ground truth BN as the constraint.

The results of ablation studies are shown in Table \ref{ablation}. As can be seen, dropping these methods brings performance degradation in terms of subjective evaluation and objective evaluation. Especially dropping both of them simultaneously leads to sifnificantly performance degradation. 
This validates the effectiveness and importance of the hierarchical CVAE modeling structure in our proposed model.



\section{CONCLUSIONS}
\label{sec:conclusions}

In this paper, we propose Accent-VITS, a VITS-based end-to-end model with a hierarchical CVAE structure for accent transfer. The hierarchical CVAE respectively models accent pronunciation information with the constraint of BN and acoustic features with the constraint of mel-spectrum. 
Experiments on professional Mandarin data and accent data show that Accent-VITS significantly outperforms the Text2BN2Mel+Neural-Vocoder three-stage approach and the VITS-DAT approach. 

%
%
%
\bibliographystyle{splncs04}
\bibliography{refs}
%




\end{document}